\documentstyle[prb,aps,psfig,twocolumn]{revtex}

\begin{document}
\twocolumn[\hsize\textwidth\columnwidth\hsize\csname
@twocolumnfalse\endcsname

\title
{Interplay between orbital ordering and lattice
distortions in LaMnO$_3$, YVO$_3$, and YTiO$_3$
}
\author{T. Mizokawa, D. I. Khomskii, and G. A. Sawatzky}
\address
{Solid State Physics Laboratory,
Materials Science Centre,
University of Groningen,
Nijenborgh 4, 9747 AG Groningen,
The Netherlands}
\date{\today}
\maketitle

\begin{abstract}
We have studied the interplay between orbital ordering,
Jahn-Teller and GdFeO$_3$-type lattice distortions in
perovskite-type transition-metal oxides
using model Hartree-Fock calculations.
It has been found that 
the covalency between $A$-site cations and oxygens causes 
interaction between the Jahn-Teller
and GdFeO$_3$-type distortions.
The present calculations explain why the $d$-type Jahn-Teller 
distortion and orbital ordering compatible with it 
are realized in LaMnO$_3$, YVO$_3$ and YTiO$_3$.
\end{abstract}


]

\section{Introduction}
\begin{figure}
\psfig{figure=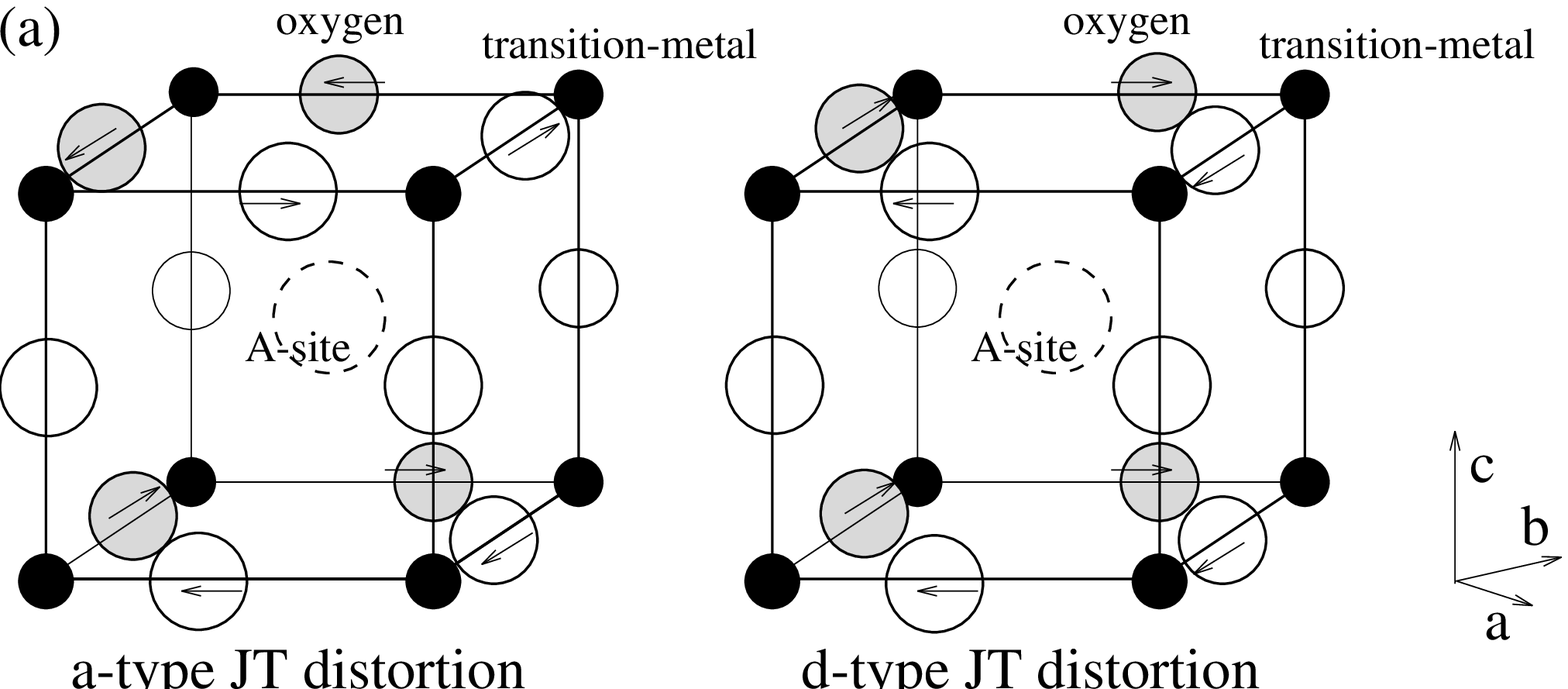,width=8cm}
\vskip 2mm
\psfig{figure=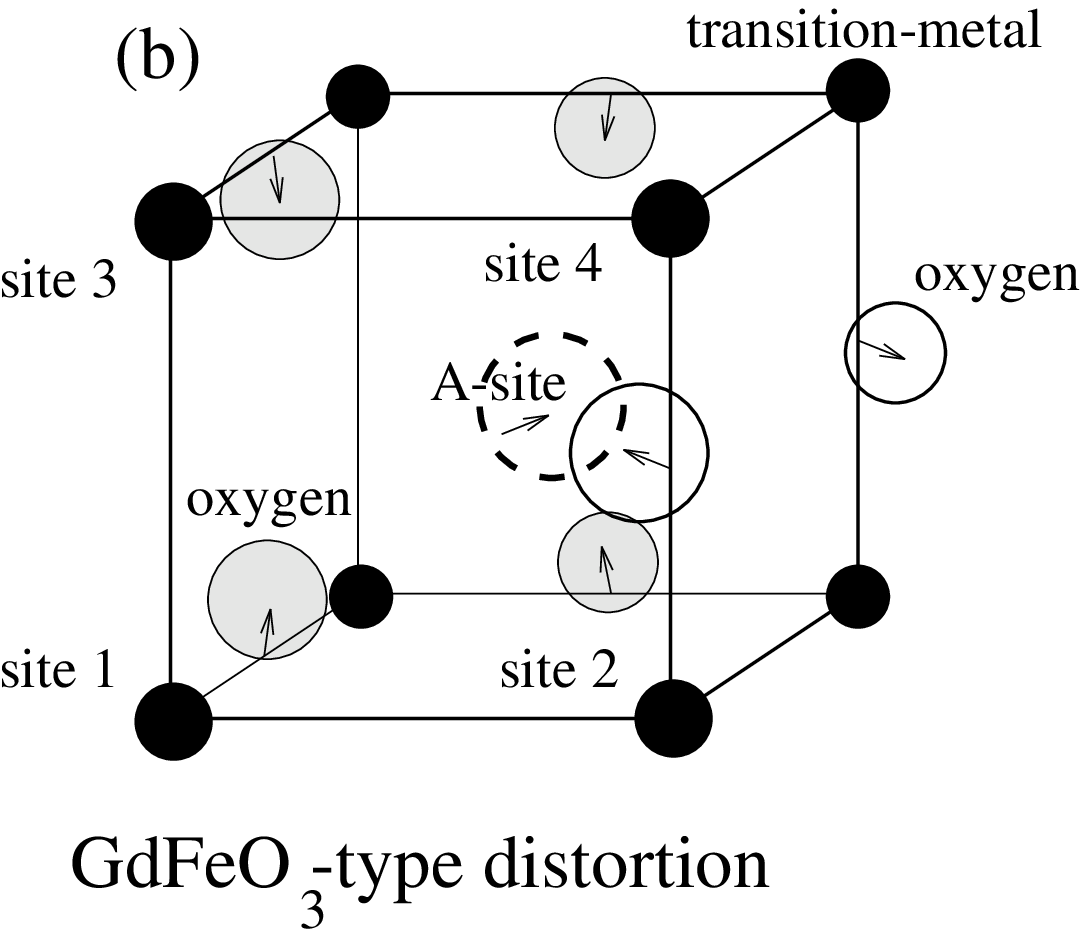,width=4.5cm}
\vskip 2mm
\caption{
Schematic drawings
for two types of Jahn-Teller distortions (a)
and for GdFeO$_3$-type distortion (b).
The arrows indicate shifts of the oxygen 
and $A$-site ions. In the GdFeO$_3$-type distortion,
the six oxygen ions nearest to the $A$-site ion
are shown.
}
\label{distortion}
\end{figure}

Orbital ordering and concomitant Jahn-Teller (JT) distortions
are observed in some perovskite-type 3$d$ transition-metal compounds
such as KCuF$_3$ and LaMnO$_3$. \cite{KK,KCuF3,LaMnO3,LaMnO3n}
In the perovskite-type lattice, there are two possible JT
distortions depending on the stacking of the elongated octahedra
along the $c$-axis as shown in Fig. \ref{distortion}(a). \cite{KCuF3}
In the $d$-type JT distortion,
the elongated axes of the octahedra are parallel along the $c$-axis.
On the other hand,
the elongated axes are rotated by 90$^\circ$ along the $c$-axis
in the $a$-type JT distortion.
While LaMnO$_3$ ($d^4$), YVO$_3$ ($d^2$) and YTiO$_3$ ($d^1$) have  
the $d$-type JT distortion, \cite{LaMnO3,YVO3,YTiO3} 
LaVO$_3$ ($d^2$) has
the $a$-type JT distortion. \cite{LaVO3}
In KCuF$_3$, both the $d$-type and the $a$-type JT distortions
are observed. \cite{KCuF3} 
Hartree-Fock calculations which consider
the hybridization between the transition-metal 3$d$ and oxygen 2$p$
orbitals predict that the orbital ordered state compatible with 
the $a$-type JT distortion is lower in energy
than that with the $d$-type JT distortion
for $d^1$ and $d^2$ systems and that the two states
are degenerate for $d^4$ and $d^9$ systems. \cite{Mizokawa}
Therefore, one cannot explain why the $d$-type JT distortion
is realized in LaMnO$_3$, YVO$_3$ and YTiO$_3$ by considering
the energy gain due to orbital ordering alone.

Perovskite-type $AB$O$_3$ compounds with relatively small $A$-site
ions undergo the GdFeO$_3$-type distortion which is caused by tilting
of $B$O$_6$ octahedra as shown in Fig. \ref{distortion}(b). 
\cite{GdFeO3}
While LaMnO$_3$, YVO$_3$, LaVO$_3$ and YTiO$_3$  
are accompanied by the GdFeO$_3$-type distortion, 
KCuF$_3$ has no GdFeO$_3$-type distortion.
In addition, the magnitude of the GdFeO$_3$-type distortion
becomes larger in going from LaVO$_3$ and LaMnO$_3$ to YVO$_3$ 
and YTiO$_3$. 
Here, one can notice that the compounds with the larger
GdFeO$_3$-type distortion tend to have the $d$-type JT distortion.
It has been pointed out by Goodenough that the covalency between
the $A$-site and oxygen ions ($A$-O covalency) is important in the 
GdFeO$_3$-type distortion. \cite{covalency}
In this paper, we have studied the relationship
between the GdFeO$_3$-type and JT distortions
considering the $A$-O covalency
and explored the reason why  
orbital ordering compatible with the $d$-type JT 
distortion are favored in LaMnO$_3$, YVO$_3$ and YTiO$_3$ 
in terms of the interaction between the two distortions.

\section{Method of calculation}
 
We have employed lattice models for the perovskite-type
structure in which the transition-metal 3$d$,
the oxygen 2$p$ and the $A$-site cation $d$ orbitals
are included. The on-site Coulomb interaction between
the transition-metal 3$d$ orbitals, which is essential
to make the system insulating and to cause orbital ordering,
are expressed using Kanamori parameters $u$, $u'$, $j$ and $j'$. 
\cite{Kanamori} The charge-transfer energy $\Delta$ is defined
by $\epsilon^{0}_{d}-\epsilon_{p}+nU$, where 
$\epsilon^{0}_{d}$ and $\epsilon_{p}$ are the energies
of the bare transition-metal 3$d$ and oxygen 2$p$ orbitals
and $U$ (=$u$-20/9$j$) is the multiplet averaged 
$d$-$d$ Coulomb interaction energy. 
The hybridization between the transition-metal 3$d$ and 
oxygen 2$p$ orbitals is expressed by Slater-Koster parameters 
($pd\sigma$) and ($pd\pi$). The ratio ($pd\sigma$)/($pd\pi$)
is fixed at -2.16. \cite{Mat}
$\Delta$, $U$, and ($pd\sigma$) can be deduced from cluster-model
analysis of photoemission spectra. \cite{Mizokawa}
Although the error bars of these parameters
estimated from photoemission spectra are not so small 
[$\sim \pm$1 eV for $\Delta$ and $U$ and $\sim \pm$0.2 eV
for $(pd\sigma)$], the conclusions obtained in the present
calculations are not changed if the parameters 
are varied within the error bars.

In the present model, 
unoccupied $d$ orbitals of the $A$-site cation such as Y 4$d$ 
and La 5$d$ are taken into account. The hybridization term
between the oxygen 2$p$ orbitals and $A$-site cation
$d$ orbitals is expressed by $(pd\sigma)_A$ and $(pd\pi)_A$.
The ratio $(pd\sigma)_A$/$(pd\pi)_A$ is also fixed at -2.16.
The hybridization term between the oxygen 2$p$ orbitals is
given by ($pp\sigma$) and ($pp\pi$) and the ratio
($pp\sigma$)/($pp\pi$) is fixed at -4. \cite{Mat}
It is assumed that the transfer integrals
$(pd\sigma)$ and $(pd\sigma)_A$ 
are proportional to $d^{-3.5}$ and $(pp\sigma)$ is to $d^{-2}$, 
where $d$ is the bond length. \cite{Mat}
Without the JT and GdFeO$_3$-type distortions,
the bond length between two neighboring oxygens
and that between the oxygen and the $A$-site
cation are $\sqrt{2} a$, where $a$ is the bond length between
the transition-metal ion and the oxygen.
($pp\sigma$) and $(pd\sigma)_A$ for bond length of $\sqrt{2} a$
are assumed to be -0.60 and -1.0 eV, respectively.

In the GdFeO$_3$-type distortion, the four octahedra
in the unit cell are rotated by angle of $\omega$ around the axes
in the (0,1,1) plane in terms of the orthorombic unit cell.
Here, we model the GdFeO$_3$-type distortion by rotating
the octahedra around the (0,1,0)-axis or the $b$-axis 
[see Fig. \ref{distortion}(b)].
The subsequent small rotation around the $a$-axis is required  
to retain the corner-sharing network of the octahedra.
The magnitude of the GdFeO$_3$-type distortion is expressed
by the tilting angle $\omega$ around the $b$-axis. 
It is important that the $A$-site ions are shifted 
approximately along the $b$-axis to decrease the distance 
from the $A$-site ion to the three closest oxygen ions
and increase the distance to the three next closest oxygen
ions as shown in 
Fig. \ref{distortion}(b).
Here, it is assumed that the shift is along the 
($\pm$1/8,7/8,0)-direction. The magnitude of the shift
is proportional to the tilting angle and is assumed
to be $\sim$ 0.05$a$, 0.1$a$ and 0.15$a$ for the tilting
angles of 5, 10 and 15$^\circ$, respectively,
which are realistic values for the compounds studied
in the present work. \cite{LaMnO3,LaVO3} 
As for the Jahn-Teller distortion, it is assumed that  
the longest bond is by 0.1$a$ longer than the shortest bond
which is reasonable for LaMnO$_3$ and is relatively large 
for LaVO$_3$ and YTiO$_3$. \cite{LaMnO3,LaVO3,YTiO3} 

\section{Results and Discussion}

\subsection{LaMnO$_3$}

\begin{figure}
\psfig{figure=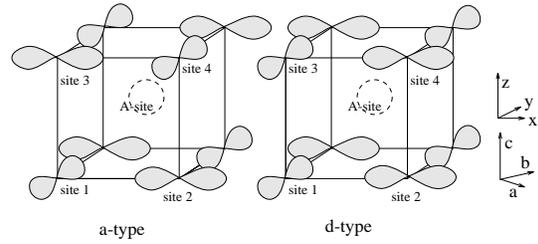,width=7cm}
\vskip 2mm
\caption{
Two types of orbital ordering for LaMnO$_3$
compatible with the $a$-type and $d$-type 
JT distortions.
}
\label{orbital}
\end{figure}

In the high-spin $d^4$ system, in which 
one of the $e_g$ orbitals is occupied at each site, 
the $A$-type antiferromagnetic (AFM) states
with $3x^2-r^2$/$3y^2-r^2$-type orbital ordering
with considerable mixture of $3z^2-r^2$
are predicted to be stable by theoretical
calculations \cite{KK,Mizokawa,Theory} 
and are studied by x-ray and neutron diffraction
measurements. \cite{LaMnO3n} Here, the $z$-direction
is along the $c$-axis. The amount of the $3z^2-r^2$
component decreases with the JT distortion. \cite{Mizokawa}
Different ways of stacking the orbitals along the $c$-axis give
two types of orbital ordering:
the one compatible with the $d$-type JT distortion
and the other with the $a$-type JT distortion.
These two types of orbital ordering are illustrated 
in Fig. \ref{orbital}. While, in the orbital ordering of the $a$-type,
the sites 1, 2, 3, and 4 are occupied by
$3y^2-r^2$, $3x^2-r^2$, $3x^2-r^2$, and $3y^2-r^2$ orbitals,
the sites 1, 2, 3, and 4 are occupied by
$3y^2-r^2$, $3x^2-r^2$, $3y^2-r^2$, and $3x^2-r^2$ orbitals
in the orbital ordering compatible with the $d$-type JT distortion.

\begin{figure}
\psfig{figure=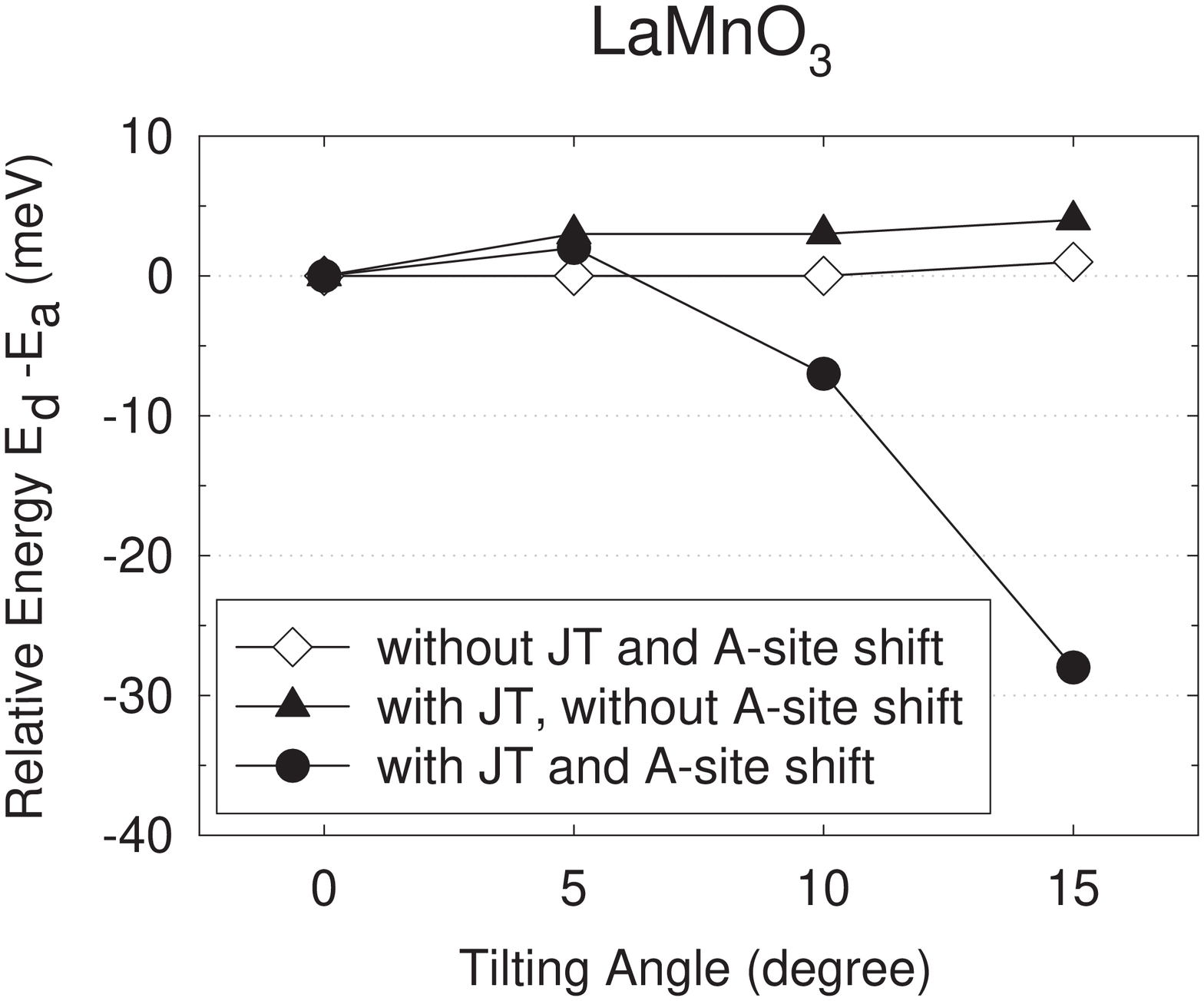,height=7cm}
\caption{
Energies per formula unit cell of the orbital ordered state
compatible with the $d$-type JT distortion
relative to that with the $a$-type JT distortion for LaMnO$_3$ 
as a function of the tilting angle, i.e., the
magnitude of the GdFeO$_3$-type distortion. 
closed circles: with the JT distortions and  
with the shift of the $A$-site cation, 
open diamonds: without the JT distortions and
without the shift of the $A$-site cation,
closed triangles: with the JT distortions and without 
the shift of the $A$-site cation.
}
\label{LaMnO3}
\end{figure}

In Fig. \ref{LaMnO3}, the energy difference between the orbital
ordered states compatible with the $d$-type and $a$-type
JT distortions is plotted as a function of 
the tilting angle of the octahedra, i.e., the magnitude of
the GdFeO$_3$-type distortion.
$\Delta$, $U$, and $(pd\sigma)$ are 4.0, 5.5, and -1.8 eV,
respectively, for LaMnO$_3$. \cite{Mizokawa}
Without the JT distortion and the shift of the $A$-site ion,
the two states are degenerate within the accuracy of the
present calculation ($\pm$ 1 meV/formula unit cell). 
This degeneracy is
lifted when the JT distortion is included.
With the JT distortion and the shift of the $A$-site ion, 
the orbital ordered state with the $d$-type JT distortion
is lower in energy than that with the $a$-type JT distortion.
If we tentatively switch off the shift of the $A$-site ion
and include only the JT distortion,
the orbital ordered state with the $a$-type JT distortion becomes 
slightly lower than that with the $d$-type JT distortion
as shown in Fig. \ref{LaMnO3}.  
Therefore, one can conclude that
the shift of the $A$-site ion 
driven by the GdFeO$_3$-type distortion
is essential to stabilize the orbital ordered state with
the $d$-type Jahn-Teller distortion. \cite{Made}

\begin{figure}
\psfig{figure=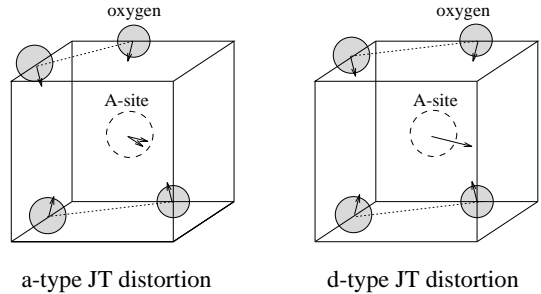,width=7cm}
\vskip 2mm
\caption{
A rough sketch of the  
shifts of the oxygen ions and the forces to 
the $A$-site ion
for the $a$-type and $d$-type JT distortions.
The oxygen shifts due to GdFeO$_3$-type distortion 
are exaggerated. Only the shifts 
and forces relevant for the present discussion are shown.
In the $a$-type JT distortion, the oxygen ions
in the upper and lower planes push the $A$-site ion
in different directions. In the $d$-type JT distortion,
the oxygen ions push the $A$-cite ion in the same direction. 
}
\label{push}
\end{figure}

The qualitative explanation of this behavior is as follows.
In the $d$-type JT distortion,
the four oxygen ions nearest to the $A$-site ion 
[shaded in Fig. \ref{distortion} (a) and (b)]
shift approximately in the same direction and, consequently, 
the system can gain the hybridization energy 
between the $A$-site and oxygen ions effectively.
On the other hand, in the $a$-type JT distortion,
the two of the four oxygen ions move in the other direction
and the energy gain due to the hybridization
is small compared to the $d$-type JT distortion. 
Another possible picture is that, in the $d$-type JT distortion,
these four oxygen ions can push the $A$-site ion in the
same direction since the JT distortion along the $c$-axis
is in phase. 
On the other hand, in the case of the $a$-type
JT distortion, the two oxygen ions in the upper plane push 
the $A$-site ion in the other direction than the two
in the lower plane as shown in Fig. \ref{push}. 
Therefore, the stronger is the
GdFeO$_3$-type distortion, the more does it stabilize 
the $d$-type Jahn-Teller distortion and 
corresponding orbital ordering.

In the charge-ordered state of Pr$_{0.5}$Ca$_{0.5}$MnO$_3$, 
the Mn$^{3+}$ and Mn$^{4+}$ sites are arranged like a checkerboard
within the $c$-plane and the same arrangement
is stacked along the $c$-axis. \cite{Jirak}
The Mn$^{3+}$ sites are accompanied by the JT distortion and 
the elongated axes are parallel along the $c$-axis just
like the $d$-type JT distortion. 
Since the $A$-sites are occupied by Pr and Ca ions
in Pr$_{0.5}$Ca$_{0.5}$MnO$_3$, we cannot simply
apply the present model calculation to it.
However, it is reasonable to speculate that 
the stacking along the $c$-axis in Pr$_{0.5}$Ca$_{0.5}$MnO$_3$
is also determined by
the interaction between the JT distortion and 
the shift of the $A$-site ion in the same way
as in LaMnO$_3$.

\subsection{YVO$_3$}

\begin{figure}
\psfig{figure=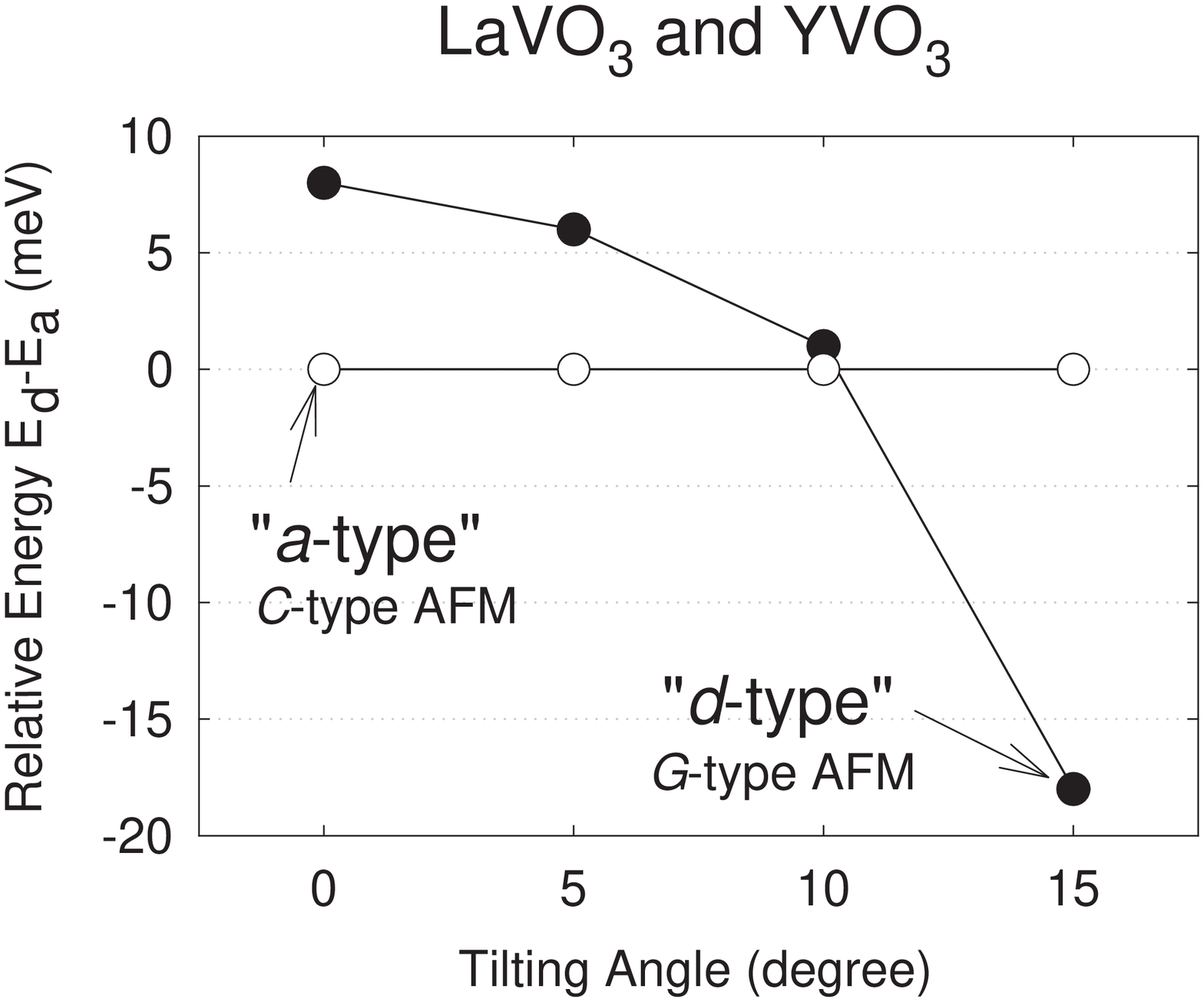,height=7cm}
\caption{
Energies per formula unit cell of the orbital ordered state
compatible with the $d$-type JT distortion (closed circles)
relative to that with the $a$-type JT distortion (open circles) 
for YVO$_3$ and LaVO$_3$ as a function of the tilting.
The shift of the $A$-site cation is included.
}
\label{YVO3}
\end{figure}

In the $d^2$ system, the $C$-type AFM state in which 
the sites 1, 2, 3, and 4 are occupied by $xy$ and $yz$,
$xy$ and $zx$, $xy$ and $zx$, and $xy$ and $yz$ orbitals 
and the $G$-type AFM state in which
the sites 1, 2, 3, and 4 are occupied by $xy$ and $yz$,
$xy$ and $zx$, $xy$ and $yz$, and $xy$ and $zx$ orbitals 
are competing.
While the $C$-type AFM state is favored by
the orbital ordering which is compatible with the $a$-type
JT distortion, the $G$-type AFM state is favored by 
the orbital ordering of the $d$-type.
The relative energy of the $G$-type AFM state with the
$d$-type JT distortion
to the $C$-type AFM state with the $a$-type JT distortion,
$E_d-E_a$, 
is plotted as a function of the tilting angle in Fig. \ref{YVO3}. 
$\Delta$, $U$, and $(pd\sigma)$ are 6.0, 4.5, and -2.2 eV,
respectively, for LaVO$_3$ and YVO$_3$. \cite{Mizokawa}
Without the GdFeO$_3$-type distortion,
the $C$-type AFM state with the $a$-type JT distortion
is lower in energy than the $G$-type AFM state, indicating that
the energy gain due to the orbital ordering is larger in the
$C$-type AFM state than in the $G$-type AFM state. 
The energy difference becomes smaller with the tilting
or the GdFeO$_3$-type distortion.
Finally, with the tilting of 15$^\circ$,
the $G$-type AFM state with the $d$-type JT distortion
becomes lower in energy than the $C$-type AFM state.
The present calculation is in good agreement with
the experimental result that the less distorted LaVO$_3$
is $C$-type AFM below 140 K \cite{LaVO3} and 
the more distorted YVO$_3$ is $G$-type AFM
below 77 K. \cite{YVO3}

\begin{figure}
\psfig{figure=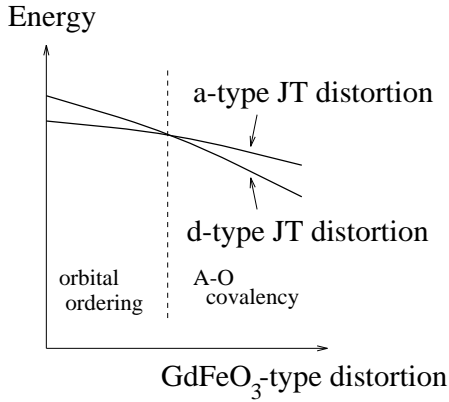,width=6cm}
\vskip 2mm
\caption{
Energy scheme of the orbital ordered states with the
$d$-type and $a$-type JT distortions
as a function of tilting or the GdFeO$_3$-type distortion.
While, for the small GdFeO$_3$-type distortion, the energy gain 
due to the orbital ordering is dominant, 
the energy gain due to the $A$-O covalency becomes relevant
for the large GdFeO$_3$-type distortion.
}
\label{YVO3s}
\end{figure}

This situation is illustrated in Fig. \ref{YVO3s}.
When the GdFeO$_3$-type distortion is large,
the interaction between the $d$-type JT distortion
and the shift of the $A$-site ion, namely, the energy gain
due to $A$-O covalency becomes dominant just like in LaMnO$_3$
and, consequently, 
the $G$-type AFM with the $d$-type JT distortion is favored.
On the other hand, when the GdFeO$_3$-type distortion is small,
the energy gain due to orbital ordering becomes dominant
and the $a$-type JT distortion and the orbital ordering 
compatible with it are realized. 
The JT distortion may be suppressed
if the system is located near the crossing point where
the $a$-type and $d$-type JT distortions are almost degenerate. 
An interesting experimental result related to this point
is that YVO$_3$ 
becomes $C$-type AFM between 77 K and 118 K.\cite{YVO3}
The present model calculation suggests that, if 
the JT distortion is switched off, the $C$-type AFM
state is favored because of the orbital ordering. \cite{Mizokawa}
Therefore, YVO$_3$ is expected to be close to the crossing point
and become $C$-type AFM when the $d$-type JT distortion
is suppressed at elevated temperature. 
In this sense, between 77 K and 118 K, YVO$_3$ 
may be an ideal orbitally ordered system 
without JT distortion.

\subsection{YTiO$_3$}

\begin{figure}
\psfig{figure=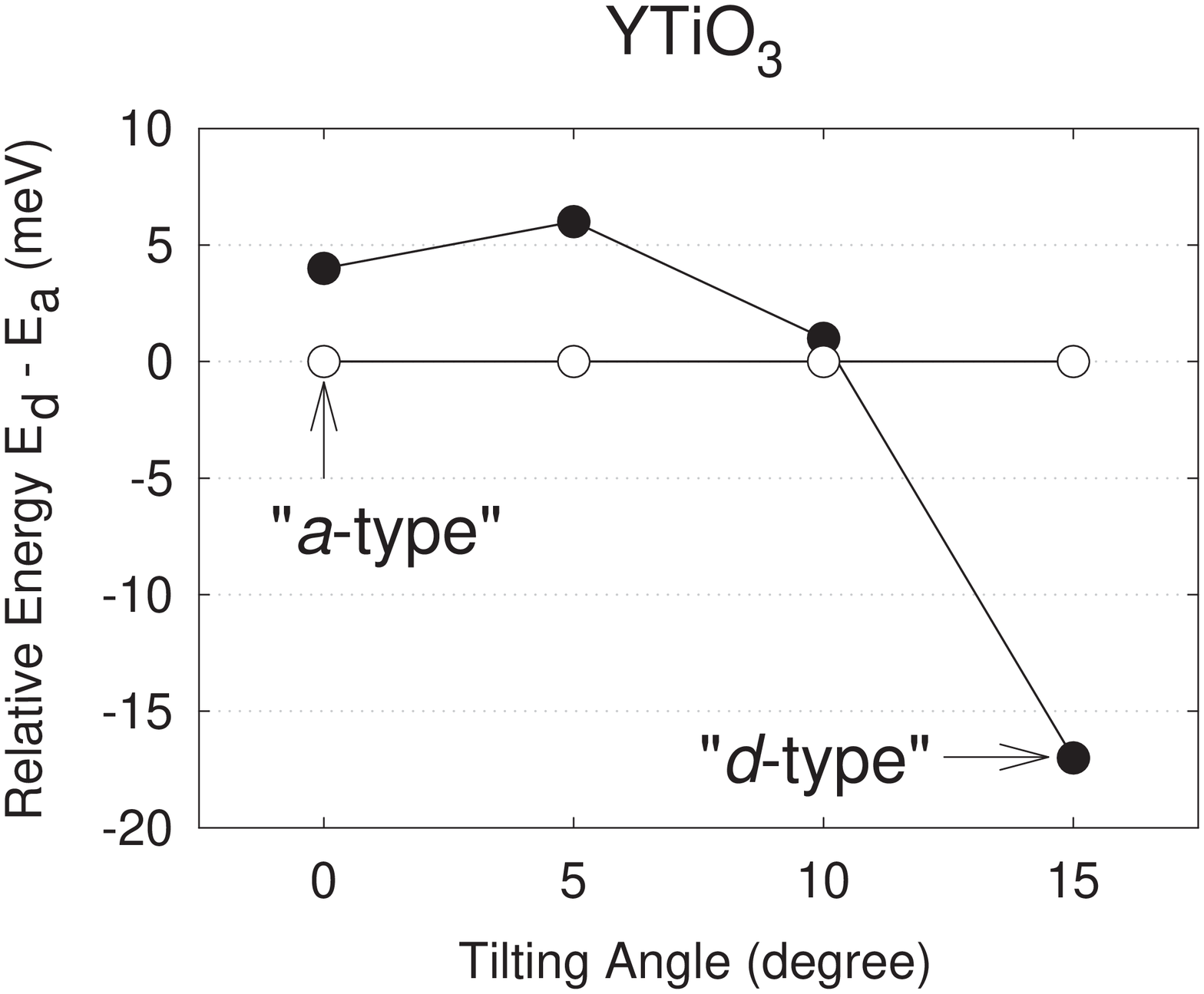,height=7cm}
\caption{
Energies per formula unit cell of the orbital ordered state
compatible with the $d$-type JT distortion (closed circles)
relative to that with the $a$-type JT distortion (open circles)
for YTiO$_3$ as a function of the tilting.
The shift of the $A$-site cation is included.
}
\label{YTiO3}
\end{figure}

For the $d^1$ system, the ferromagnetic (FM) states with 
orbital ordering are favored in the model HF calculations. 
There are two possible orbital orderings
compatible with the $a$-type and $d$-type JT distortions. 
The model HF calculation without the covalency between 
the $A$-site cation and the oxygen ion predicted that 
the orbital ordering of the $a$-type
is lower in energy. However, the recent neutron
diffraction measurement by Akimitsu {\it et al.} have shown
that the orbital ordering compatible with the $d$-type 
JT distortion is realized in the FM insulator YTiO$_3$. \cite{YTiO3}
YTiO$_3$ has the considerable GdFeO$_3$-type distortion
and the tilting angle is expected to be larger than 15$^\circ$.
In Fig. \ref{YTiO3}, the relative energy of the FM and 
orbital ordered state of the $d$-type 
to that of the $a$-type is plotted as a function
of the tilting. 
$\Delta$, $U$, and $(pd\sigma)$ are 7.0, 4.0, and -2.2 eV,
respectively, for YTiO$_3$. \cite{Mizokawa}
With the tilting of 15$^\circ$, 
the orbital ordered state of the  
$d$-type is lower in energy,
in agreement with the experimental result. 
In this state, the sites 1, 2, 3, and 4
are occupied by $c_1yz+c_2xy$, $c_1zx+c_2xy$, $c_1yz-c_2xy$,
and $c_1zx-c_2xy$ orbitals ($c_1 \sim 0.8$ and $c_2 \sim 0.6$).
However, experimentally, less distorted LaTiO$_3$ has no or
very small JT distortion and has a $G$-type AFM state.
\cite{LaTiO3}
The present calculation cannot explain why the $G$-type
AFM state can be stable compared to the FM state  
in LaTiO$_3$.

\section{Conclusion}

In conclusion, we have studied the relationship
between orbital ordering and the JT and GdFeO$_3$-type
lattice distortions. It has been found that the covalency 
between the $A$-site cations and oxygen
makes the $d$-type JT distortion (same orbitals 
along the $c$-direction) lower in energy
than the $a$-type JT distortion (alternating orbitals
along the $c$-direction) in the presence of 
the large GdFeO$_3$-type distortion. As a result,
the orbital ordered states compatible with the $d$-type
JT distortion are favored in LaMnO$_3$, YVO$_3$, and
YTiO$_3$ which have the relatively large 
GdFeO$_3$-type distortion. On the other hand,
in less distorted LaVO$_3$, the orbital ordering
compatible with the $a$-type JT distortion is favored
because of the pure superexchange effect. 

\section*{Acknowledgment}

The authors would like to thank useful discussions
with K. Tomimoto, J. Akimitsu, Y. Ren and P. M. Woodward.
This work was supported by
the Nederlands foundation for Fundamental Research
of Matter (FOM).

\begin{references}
\bibitem{Made}
We have evaluated the Madelung energy of the model
perovskite lattices with the JT and GdFeO$_3$-type distortions
and have found that the $a$-type JT distortion is slightly
lower in energy than the $d$-type. This is opposite to
the effect of the covalency between the $A$-site cations
and oxygens.
\end{references}
\end{document}